\documentstyle[12pt]{article}

\def\lsim{\mathrel{\rlap{\lower3pt\hbox{\hskip0pt$\sim$}}
    \raise1pt\hbox{$<$}}}         
\def\gsim{\mathrel{\rlap{\lower4pt\hbox{\hskip1pt$\sim$}}
    \raise1pt\hbox{$>$}}}         
\def\beq{\begin{equation}}   \def\eeq{\end{equation}}

\begin{document}
\begin{titlepage}
\renewcommand{\thefootnote}{\fnsymbol{footnote}}

\begin{center} \Large
{\bf Theoretical Physics Institute}\\
{\bf University of Minnesota}\vspace{0.2cm}
\end{center}
\begin{flushright}

TPI-MINN-98/34-T\\
UMN-TH-1723-98\\
hep-th/9809184\\

\end{flushright}

\vspace{2cm}
\begin{center}

   {\LARGE Domain Walls and  Decay Rate of the Excited Vacua
in the Large $N$ Yang-Mills Theory}

\vspace{0.5cm}
\end{center}
\begin{center} {\Large M. Shifman}\\

\vspace{0.2cm}
{\it  Theoretical Physics Institute, Univ. of Minnesota,
Minneapolis, MN 55455}
\end{center}

\vspace*{.4cm}
\begin{abstract}
In the (non-supersymmetric) Yang-Mills theory in the large $N$ limit
there exists an infinite set of non-degenerate ``vacua''.
 The distinct vacua are separated by  domain walls
 whose tension determines the decay rate of the false vacua. I discuss 
the phenomenon from a field-theoretic 
point of view, starting from supersymmetric gluodynamics and then
breaking supersymmetry  by introducing 
 a gluino mass. By combining
previously known results, the decay rate of the
excited vacua is estimated,
$\Gamma \sim \exp (-{\rm const}\times N^4)$. The fourth power of 
$N$ in the exponent
is a consequence of the fact that the wall tension 
is proportional to  $N$.

\end{abstract}

\end{titlepage}

The $\theta$ dependence of the pure Yang-Mills theories
in the strong coupling regime is being  investigated
for a long time. A qualitative picture gradually emerged 
explaining various observations regarding the $\theta$
dependence of the vacuum energy $E$. This picture
 (for a very clear summary see Sec. 1 in Ref.
\cite{EW1}) predicts the existence of a set of states
-- only one of them is the true vacuum while all others are
``excited vacua''
-- intertwined together in the process of the
$\theta$ evolution. Each time $\theta$ crosses
$\pi$, $3\pi$, and so on  the levels cross and change their relative 
roles:
one of the excited vacua becomes the true one and {\em vice versa}.
Then, $E(\theta )$ is a multibranch function of $\theta$
of the type
$$
E(\theta )=  N^2\,  {\rm min}_k\,  F\left( \frac{\theta 
+ 2\pi k }{N}\right)
$$
where $N$ is the number of colors and $F(x)$
is some $N$ independent function. This vacuum structure can be 
proven
in softly broken
supersymmetric (SUSY)  theories \cite{2,3}.
Recently it was derived
\cite{EW1}  in the context of ideas connected with the
correspondence between the conformal gauge field theory
and quantum gravity on anti-de-Sitter space \cite{M}.
Maldacena's duality is believed to give rise to a large $N$ gauge 
theory
belonging to the same universality class as QCD, starting from a string 
theory.
It was shown \cite{EW1},
that for every $\theta$, there is a set of infinitely many vacua
stable at $N=\infty$. The true vacuum is obtained by minimizing 
energy over this set.
Cusps occur at $\theta = \pi (2k+1)$ where $k$ is an integer.
At these points an additional two-fold
degeneracy emerges. The adjacent vacua from the above  set
are separated by domain walls which can be described
in terms of wrapped six-branes \cite{EW1}.

Here I describe how these qualitative results
regarding the structure of the QCD vacuum
are naturally obtained in the field theory {\em per se},
and calculate the life time of the ``false''
vacua, which turns out to be proportional to
$\exp (- C\, N^4 )$. The constant $C$ can be found up to a numerical 
factor 
of order unity; the uncertainty is due to extrapolation from the 
supersymmetric to non-SUSY limit.
The method is based on this extrapolation
and on the fact that in the SUSY limit both,
the vacuum structure and the domain wall tension, are exactly
known \cite{KS,DS,KSS}. I start from SUSY gluodynamics
and, in order to break supersymmetry, introduce a mass term
$m_g$ to the gluino fields. At $m_g\ll\Lambda$ 
(where $\Lambda$ is the dynamical mass scale)
calculations are exact. As $m_g$ grows and, eventually,
crosses $\Lambda$ the gluinos decouple, and one recovers pure 
Yang-Mills 
theory. Extrapolating the small $m_g$ results to
$m_g\sim \Lambda$ yields the overall structure of the
pure Yang-Mills theory we are interested in, allowing one to 
fully establish the $N$ dependencies. In fact, this approach has been 
already applied previously \cite{2,3}. A new element which I add is
 combining it with the $N$ counting. A surprising finding
is the existence  of the stationary
domain walls in the (non-supersymmetric) Yang-Mills theory
in the limit $N\to\infty$, whose tension can be evaluated.
These domain walls occur as the boundaries
separating the distinct stable vacua
from the intertwined set.

The vacuum structure in SUSY gluodynamics
is very simple \cite{EW2,SV1,KS}. We have $N$ degenerate chirally
 asymmetric vacua, labeled by the value of the
gluino condensate
\beq
\langle {\rm Tr} \lambda\lambda\rangle
= N\, \Lambda^3\, \exp\left[\frac{i (2\pi k +\theta )}{N}\right]
\, , \,\,\,k = 0, 1, ... , N-1\, ,
\eeq 
plus a possible chirally symmetric vacuum at $\langle 
{\rm Tr} \lambda\lambda\rangle
=0$. The chirally asymmetric vacua form a family of $N$ states
intertwined in the process of the $\theta$ evolution.
At $\theta =\pi$, $3\pi , ...$ the vacuum restructuring takes place, so 
that
the physical $2\pi$ periodicity in $\theta$ is maintained.
The chirally symmetric vacuum at
$\langle {\rm Tr} \lambda\lambda\rangle = 0$ plays no role
in this process, and will be disregarded hereafter.

With the exact supersymmetry, all $N$ vacua from the
above family have the vanishing energy density and are physically 
equivalent.
If one considers two distinct vacua separated ``geographically'',
the border between them is a domain wall discussed
in \cite{DS,KSS,CS}. If the wall is BPS saturated, then the tension of 
the wall separating two adjacent vacua is
\beq
\varepsilon = \frac{N}{8\pi^2}\,  \langle {\rm Tr} 
\lambda\lambda\rangle_0
\, \left| \exp \frac{2\pi i k}{N} - \exp \frac{2\pi i (k+1)}{N}
\right|\, ,
\eeq
where $\theta$ is set equal to zero, and 
$
\langle {\rm Tr} \lambda\lambda\rangle_0
$
is the gluino condensate at $k=0$. Since $
\langle {\rm Tr} \lambda\lambda\rangle_0
$ scales as $N$, in the large $N$ limit
the tension of the saturated wall is
\beq
\varepsilon = \frac{N}{4\pi}\, \Lambda^3\,.
\label{ndepep}
\eeq 
It is important to note that
the asymptotic behavior is linear in $N$. Even in the unlikely case 
the wall
is not saturated \footnote{The issue whether or not
the walls interpolating between the adjacent vacua (the so called 
complex walls) are BPS saturated
is being debated \cite{Smilga}. For $N=2$ and 3 the amended 
\cite{KS} Veneziano-Yankielowicz  Lagrangian \cite{VY} exhibits no 
complex  walls at all. For large $N$ the walls 
separating the adjacent vacua must exist. Arguments were given
\cite{KKS} that the  Veneziano-Yankielowicz
Lagrangian is inappropriate for the explorations of the complex walls.
To see whether or not the BPS saturated walls are present,   cusps
inherent to this Lagrangian must be smoothed out, see e.g. 
\cite{Gabad}. }
its tension must differ from Eq. (\ref{ndepep})
only by a numerical factor which does not alter the $N$ dependence
of $\varepsilon$.

Now, what happens if one adds to the Lagrangian of SUSY 
gluodynamics
a soft SUSY-breaking term?

The gluino mass term has the form
\beq
\Delta {\cal L}_m = \frac{m_g}{g^2}
\langle {\rm Tr} \lambda\lambda \rangle + \,\, \mbox{H.c.}
\eeq
To begin with we assume that $m_g/g^2\ll\Lambda$. Now SUSY is 
broken, and with it
is gone the degeneracy of $N$ vacua of supersymmetric 
gluodynamics.
To first order in $m_g$ the energy density of the $k$-th vacuum 
becomes
\beq
E_k =- {\rm Re}\, \frac{m_g}{g^2}
\, \langle{\rm Tr} \lambda\lambda \rangle_k\, =
-\left( 2\cos\frac{2\pi k}{N}\right)\, \left(
 \frac{m_g}{g^2}\right)\, N\Lambda^3\,.
\label{ss}
\eeq
I assume $m_g/g^2$ to be real and positive. (This can be always 
achieved by adjusting $\theta$ appropriately.) Note that the 
combination
$m_g/g^2$ is renormalization-group invariant to leading order,
 and scales as $N$. The combination renormalization-group
 invariant to all orders can also be found \cite{HS},
$$
\frac{m_g}{g^2} - m_g\frac{N}{8\pi^2}\, .
$$
For our purposes it is sufficient to limit
ourselves to the leading order.

Generically, all vacua are shifted from zero by $\Delta E \sim N^2$,
in full accordance with the general expectations regarding the 
vacuum energy 
in the non-supersymmetric gauge theories.
The true vacuum corresponds to $k=0$. The states at 
$k\neq 0$ have a higher energy density. The spectrum of the states
corresponding to  Eq. (\ref{ss})
consists of two distinct parts (call them the first and the second part, 
respectively).
For $k$ that does not scale with $N$ the argument of the cosine is 
small,
and the level splitting between the neighboring vacua is
\beq
\Delta E \sim 8\pi^2 \left( k+\frac{1}{2}\right) \, \frac{m_g}{Ng^2}\,   
\Lambda^3\sim N^0\, .
\label{deltae}
\eeq
As is seen from this expression, 
for higher $k$ the level splittings grow
and become of order $N$ when $k$ becomes proportional to $N$.
This is the maximal dependence of the level splittings on $N$.
For $k\sim N$ Eq. (\ref{deltae}) is not valid since it was obtained by 
expanding Eq. (\ref{ss}). One can see directly from  Eq. (\ref{ss}) that at
$k\sim N$ the energy splittings $\Delta E \sim N$. 
Note that at $N=\infty $ the number of states belonging to the 
 part of the vacuum family
with the level splittings of order $N^0$ (i.e. the first part of the 
spectrum) 
is infinite by itself. The fate of the vacua from this part of the 
spectrum, on the one hand, and the 
higher-lying states (from the second part), on the other hand,  is 
different. The height of the barrier in the functional space, separating 
the adjacent vacua is of order $N$
(see Eq. (\ref{ndepep})). It is determined by the wall tension. One 
should keep in mind that the wall width $\sim \Lambda^{-1}\sim 
N^0$. Although so far the wall tension was obtained in
SUSY gluodynamics, the gluino mass term does
not affect it as long as $m_g\ll \Lambda$.
Even at $m_g\gsim \Lambda$ the walls, interpolating between those 
vacua that belong to  the first part
of the spectrum,  persist as static objects,
 and their tension changes only by order unity.
 The $N$ dependence of $\varepsilon$  remains intact.
Therefore, the vacua from the first part of the spectrum are stable
in spite of the fact that they are non-degenerate. Below we will 
evaluate their decay rate to be $\exp (-C \, N^4)$.

As for the vacua from the second part of the spectrum, at $k\propto N$ 
they may 
disappear at all  as local minima in the functional space.
Or, else,
some of them may survive as shallow minima.
In any case, they disappear as stationary states,
and the walls interpolating between these former vacua,
even if they survive as shallow minima,
are not static objects, they tend to ``decay''.  Needless to say that for 
such walls the estimate of their tension from Eq. (\ref{ndepep}) would 
be wrong. 

If the decay rates of the vacua from the first part of the family tend 
to zero at $N\to\infty$ as  $\exp (-C \, N^4)$, the decay rates of the 
states from the second part are either of order unity at $k\sim N$
or vanish slower than $\exp (-C \, N^4)$ if $k$ scales as $N^\sigma$ 
with $\sigma < 1$. 

Now we estimate the decay rate of the stable vacua.
The false ones decay into the true vacuum
through the bubble formation. 
The quasiclassical theory of these decays is well-developed
\cite{VKO}, it is applicable if the radius of the critical bubble is
 large, much larger than the wall width. In our case, the radius
of the critical bubble is proportional to  $N$ (this is the radius 
corresponding to a balance between the volume energy gained
and the surface energy lost), while the wall width is $N$ 
independent. Therefore, at large $N$ the quasiclassical theory
is valid.  The general result of this theory is
  \beq
 \label{rateVol}
 \Gamma \ \propto \ \exp\left\{ - \frac {27}2 \pi^2 \frac 
{\varepsilon^4}{(\Delta { E})^3}
\right\}\, ,
  \eeq
where $\Delta { E}$ is the difference of the vacuum energy 
densities in the false and
true vacua, and $\varepsilon$ is the surface energy density of the 
domain 
wall.  With our values of $\varepsilon$ and $\Delta E$
we get \footnote{For $N=2$ a similar calculation has been carried out 
in Ref. \cite{KSS}.}
\beq
\Gamma\sim \exp\left\{-\left| \frac{{\rm 
Tr}\langle\lambda\lambda\rangle_0}{[m/(Ng^2)]^3}
\right|\,  \frac{N^3}{\left( k +\frac{1}{2}\right)^3}\, 
\frac{3^3}{2^{18}\pi^8}
\right\}\, . 
\eeq
The result for the exponent is rigorously valid for $m_g\ll\Lambda$
and $k\ll N$.
I will now extrapolate it to the point of the gaugino decpoupling,
i.e. $m_g/(Ng^2) \sim \Lambda$ (still assuming that $k\ll N$). Then
\beq
\Gamma\sim \exp\left\{-  \eta   \frac{3^3}{2^{18}\pi^8} \, 
\frac{N^4}{\left( k +\frac{1}{2}\right)^3}\right\}\, , 
\label{fe}
\eeq
where a dimensionless coefficient $\eta$ is introduced to take 
account of the uncertainty of the extrapolation, $\eta \sim 10^0$.
This coefficient is purely numerical, it is $N$ independent.

Equation (\ref{fe}) presents an estimate of the false vacuum decay 
rate to its neighbor in the large $N$ (non-supersymmetric)
Yang-Mills theory. Even though we know the exponent
only by an order of magnitude, the presence of a very strong 
numerical suppression of the exponent seems evident. If so, our 
derivations are practically applicable only to very large $N\gsim	 
100$
even at $k\sim 1$.

It would be very interesting to check how both conclusions --
the $N^4$ functional dependence of $\ln \Gamma$ and a numerical 
suppression of the coefficient in front of $N^4$ --
appear directly within the Maldacena-Witten approach. 
Narrow quasistable excited vacua were detected recently
within  an effective Lagrangian approach in Ref. \cite{FHZ}.
Although some aspects in this consideration remain questionable and 
require further clarification 
it seems worth trying to apply the method to check whether some of 
the excited vacua survive at low $N$, and, if so,  to estimate their 
decay rate and possible phenomenological manifestations. 

One of potentially important  points is the lattice calculations. Since 
they are always done in  finite volume, which reduces the 
field-theoretic system to quantum-mechanical,  all vacua
contribute to the correlation functions calculated
in the lattice Yang-Mills theory, generally speaking.  This might lead
to a contamination of the lattice results by false vacua.
Certainly, practically all calculations are done at
$N=2$ or 3. At such low values of $N$ the false vacua
may not exist as local minima in the functional space, or may be so 
shallow, that  there is no barrier separating them
from the true one.  In this case they do not affect
determination of the physically measurable quantities
(such as the particle masses and coupling constants)
from the finite-volume lattice results. 

In summary, starting from supersymmetric gluodynamics
and extrapolating in the gluino mass one can argue
that an infinite set of the stable vacua exist in the large $N$ 
non-supersymmetric Yang-Mills theory.  Static domain walls 
interpolate between these vacua.
The decay rate of the false vacua is given by the formula (\ref{fe}).

\vspace{0.5cm}

Useful discussion with Y. Oz and A. Pasquinucci and 
correspondence with A. Smilga are
gratefully acknowledged. I would  like to thank the CERN Theory 
Division,
where this work was done, for kind hospitality. This work was 
supported
in part by DOE under  the grant number
DE-FG02-94ER40823.

\end{document}